# MID-INFRARED IMAGING OF YOUNG STELLAR OBJECTS

**Michael C. Liu, James R. Graham,[1]**
DEPARTMENT OF ASTRONOMY, UNIVERSITY OF CALIFORNIA, BERKELEY, CA 94720
email: [mliu,jgraham]@astro.berkeley.edu

**A. M. Ghez,[2]**
DEPARTMENT OF ASTRONOMY, UNIVERSITY OF CALIFORNIA, LOS ANGELES CA 90024

**M. Meixner,**
DEPARTMENT OF ASTRONOMY, UNIVERSITY OF ILLINOIS, URBANA-CHAMPAIGN, IL 61801

**C. J. Skinner,[3] Eric Keto, Roger Ball,**
INSTITUTE OF GEOPHYSICS AND PLANETARY PHYSICS,
LAWRENCE LIVERMORE NATIONAL LABORATORY, LIVERMORE, CA 94551

**J. F. Arens, and J. G. Jernigan**
SPACE SCIENCES LABORATORY, UNIVERSITY OF CALIFORNIA, BERKELEY CA 94720

## ABSTRACT

We present arcsecond resolution mid-infrared (8–13 $\mu$m) images and photometry of four young stellar objects (YSOs) — L1551-IRS5, HL Tau, AS 205, and AS 209 (V1121 Oph) — taken with the Berkeley Mid-Infrared Camera. For AS 205, a known T Tauri binary, we also present near-infrared JHK images and HKL′ speckle imaging data. All three single stars are unresolved in our mid-IR images, consistent with current models of the circumstellar material associated with these objects.

Our data is the first to resolve in the mid-IR both components of the close binary AS 205 (projected separation ∼1.3″ (210 A.U.)). Both stars are classical T Tauri stars and possess the 9.7 $\mu$m silicate feature in emission. AS 205 N is the IR brighter star in our data while published observations find it to be the optically fainter star. Assuming the IR excesses of both components arise from circumstellar disks, we find the emitting regions (the inner few A.U.) of the disks to be optically thick in the mid-IR. Pre-main sequence evolutionary models suggest the AS 205 system is non-coeval; we discuss possible explanations for this result and comment on the evolutionary status of this young binary.

All of our objects, except perhaps AS 205 South, exhibit changes in their mid-IR flux in measurements separated by intervals of days up to many years; the variations range from 30–300%. For the classical T Tauri stars AS 205 North and AS 209, the magnitude of the changes seems to discount the possibility the mid-IR variations have the same origin as the optical and near-IR variability of T Tauri stars, namely accretion-related features on or near the stellar photosphere. We speculate that the cause of the variability lies in the accretion disks of these objects; the data suggest disk accretion rate fluctuations of nearly an order of magnitude. The existence of large mid-IR variability argues that simultaneous multiwavelength observations are needed for a proper analysis of YSO spectral energy distributions.

*Subject headings:* stars: pre-main sequence — stars: binaries: close, visual — stars: circumstellar material – infrared: stars – stars: individual (L1551-IRS5, HL Tau, AS 205, AS 209)

[1] Alfred P. Sloan Fellow
[2] Hubble Fellow
[3] Present address: Space Telescope Science Institute, Baltimore, MD 21218

# 1 Introduction

Insight into the process of star formation has blossomed in the past decade with the advent of long wavelength imaging technology. Young stellar objects (YSOs) are associated with large quantities of circumstellar dust and consequently emit most strongly at infrared and millimeter wavelengths. High spatial resolution observations at these wavelengths can now resolve the details of circumstellar structures and detect close embedded companions to YSOs.

Mid-IR (8–13 $\mu$m) observations can search for extended emission from warm ($T \sim 300$ K) dust around YSOs. Because the effects of atmospheric seeing decrease at longer wavelengths, diffraction-limited imaging can be achieved in the mid-IR. Subarcsecond angular resolution is possible at 3-meter telescopes, corresponding to $\sim$100 A.U. for objects in the nearest star forming regions. Such resolution represents a considerable improvement over *IRAS* data and ground-based aperture photometry. Although current theory predicts YSOs to be unresolved at this resolution, the physical conditions and distribution of circumstellar material on such scales has barely been explored. Some sources with circumstellar material in fact are extended in the mid-IR, *e.g.*, the pre-main sequence Ae star WL 16 (Moore *et al.* 1995) and the main-sequence star $\beta$ Pic (Lagage & Pantin 1994). Further mid-IR observations provide important tests of our understanding of the material around young stars.

Long wavelength imaging with high resolution is also an important tool for studying young binaries. Binaries are common in both the main sequence and pre-main sequence (PMS) population (Duquennoy & Mayor 1991; Ghez, Neugebauer, & Matthews 1993), and the relationship between the binary fraction in these two populations is an important clue in understanding binary formation (Mathieu 1994). Many PMS secondaries are very cool, *e.g.*, T Tau South (Ghez *et al.* 1991); mid-IR surveys can potentially detect companions too deeply-embedded for optical and near-IR surveys. Furthermore, only imaging can disentangle the mid-IR emission from close binaries and allow us to construct accurate spectral energy distributions (SEDs) for both components. Studies of the SEDs of single YSOs have proven useful in deducing the properties and distribution of the circumstellar material. For instance, dips in the mid-IR region of some SEDs may signal the presence of gaps in circumstellar disks opened by planet formation (Marsh & Mahoney 1992; *cf.* Boss & Yorke 1993). By extracting the SEDs of the components of young close binaries, we can study the circumstellar and circumbinary material and search for possible evidence of binary star-disk interactions, *e.g.*, as in UZ Tau (Ghez *et al.* 1994) and GW Ori (Mathieu *et al.* 1991).

We have imaged four YSOs at mid-IR wavelengths with arcsecond resolution in order to study the spatial extent of their circumstellar material, to search for unresolved companions, and to study the IR excesses in the components of young binaries. Our sample consists of three single stars which have possible companions or extended structure at other wavelengths, L1551-IRS5, HL Tau, and AS 209, and a known binary system, AS 205.

# 2 The Objects

Table 1 gives basic information for our four sources (Herbig & Bell 1988 and references therein, except where indicated), and we briefly describe each one below.

## 2.1 L1551-IRS5

This object in the Lynds 1551 dark cloud was one of the first known protostars and is surely one of the most extensively studied YSOs (Staude & Elsässer 1993; Strom *et al.* 1985, and references therein). A highly obscured Class I YSO ($\lambda F_\lambda$ rising with $\lambda$ from 1–100 $\mu$m; Lada 1991), its luminosity of $\sim$30 $L_\odot$ makes it the most luminous protostar in the Taurus star forming region. A remarkable outflow seen in optical, infrared, millimeter, and radio line emission emerges from the compact IR source on arcsecond to arcminute scales. Subarcsecond structures suggestive of a disk are seen at submillimeter and radio wavelengths (Lay *et al.* 1994; Rodríguez *et al.* 1986). Similarly, near-IR imaging shows a $3'' \times 2''$ structure, believed to be due to starlight scattering from the surface of a disk (Moneti *et al.* 1988; Hodapp *et al.* 1985; *cf.* Whitney & Hartmann 1993). Extended structures on $\sim$10$''$ scales are suggested by observations in the far-IR (50–100 $\mu$m) and submillimeter (Butner *et al.* 1991; Ladd *et al.* 1995).

## 2.2 HL Tau

HL Tau is a $\sim$4 $L_\odot$ pre-main sequence star in the L1551 cloud. It possesses a disk of molecular gas oriented perpendicular to the optical jet (Sargent & Beckwith 1991 and references therein; Lay *et al.* 1994). IR tomographic imaging reveals a butterfly-shaped reflection nebula, well-aligned with the disk (Beckwith *et al.* 1989). Radio continuum maps at 1.3 cm and 3.6 cm show flattened arcsecond-sized structures whose orientations do not correspond with those of the molecular disk or optical jet (Rodríguez *et al.* 1994). Recent *Hubble Space Telescope* images find all the optical emission is scattered light (Stapelfeldt *et al.* 1995); Beckwith & Birk (1995) and Weintraub, Kastner, & Whitney (1995) also find the near-IR emission is dominated by scattering. These results suggest the source is highly embedded and therefore quite young.

## 2.3 AS 205

Located in the Upper Scorpius Association, AS 205 lies a few degrees northwest of the $\rho$ Oph dark cloud. Within 30$'$ there are a number of *IRAS* sources and X-ray emitting T Tauri stars (Carballo, Wesselius, & Whittet 1992; Walter 1986). Merrill & Burwell (1951) first recognized AS 205 as an emission-line star, and Herbig & Rao (1972) identified it as a close binary. Both components are classical T Tauri stars (CTTS), showing strong H$\alpha$, H$\beta$, He I, [O I], and Ca II emission lines (Cohen & Kuhi 1979; Herbig 1994) and possessing infrared excesses (§5.2.2). The combined-light spectrum has the Li I 6707 Å absorption line, an indication of the system's youth (Herbig 1977). Felli *et al.* (1982) detected strong 6 cm radio emission from the star, and Walter & Kuhi (1984) observed a large, slow x-ray flare, suggesting intermittant stellar activity. Evidence for variability also comes from comparison of published veiling measurements (Basri & Bathala 1990; Valenti, Basri, & Johns 1993), which suggest changes in accretion-related photospheric hot spots or boundary layers (Basri & Bathala; Hartigan *et al.* 1991).

## 2.4   AS 209 (V1121 Oph)

AS 209 lies in a region with a few dark clouds, *IRAS* sources, and X-ray emitting PMS stars (de Geus, Bronfman, & Thaddeus 1990; Carballo *et al.* 1992; Walter *et al.* 1994), east of the $\rho$ Oph cloud. The star possesses typical signatures of a classical T Tauri star: H$\alpha$, H$\beta$, and Ca II emission lines along with strong UV and IR excesses (Merrill & Burwell 1951; Hamann & Persson 1992; Valenti *et al.* 1993). It also shows Li I absorption in its spectrum (Herbig 1977). One-dimensional near-IR speckle observations by Weintraub & Shelton (1989) found the source to be extended, though they could not conclude whether this was due to binarity or extended structure. Subsequent two-dimensional near-IR speckle work by Ghez *et al.* (1993) discounted any bright extended emission or any binary companion with $\Delta K \leq 4$ at a separation of 0.1–0.8″.

# 3   Observations and Data Analysis

## 3.1   Mid-Infrared Imaging

We obtained mid-infrared observations in May 1990, September 1990, and May 1991 with the Berkeley Mid-Infrared Camera (Arens *et al.* 1987; Keto *et al.* 1992) at the 3-meter Infrared Telescope Facility (IRTF) located on Mauna Kea, Hawaii.[4] The camera used a 10 × 64 pixel, Ga-doped Si array. Observations were made with a 10% circular variable filter ($\Delta\lambda/\lambda = 0.10$) centered at wavelengths from 8.5–12.5 $\mu$m. The plate scale was 0.39″/pixel, and the long axis of the array was oriented north-south. We employed standard chop-nod (double-differencing) procedures to remove the thermal contribution from the sky and telescope (Ball *et al.* 1992; Meixner *et al.* 1993). Flat-fielding was done using dark-subtracted sky frames. To minimize the effects of windshake and poor telescope tracking, individual images were expanded onto a 0.2 pixel grid using bilinear interpolation and then registered and combined using a least-squares fitting routine (Meixner 1993). Figure 1 shows our AS 205 images from May 1991; all other targets were unresolved.

Mid-IR flux calibration was determined for each source by observing a standard star, either $\alpha$ Her, $\alpha$ Tau, or $\phi$ Oph, close in time and in airmass (<0.2 airmass difference). Narrow-band magnitudes were obtained from Hanner *et al.* (1984) and absolute calibrations from Tokunaga (1988) and an unpublished standards list for the mid-IR spectrograph CGS3 at the UK Infrared Telescope (Skinner 1993). We estimate a 10% error in our mid-IR photometry based on past experience with the photometric accuracy of similar data (Meixner *et al.* 1993) as well as possible low-level variability in $\alpha$ Her (Hanner *et al.*).

Table 2 summarizes our imaging observations including photometry of the sources and measurements of the 1$\sigma$ noise and angular resolution of our final mosaics. The resolution was estimated using the full width at half maximum (FWHM) of azimuthally-averaged radial profiles of the standard stars.

---

[4]The IRTF is operated by the University of Hawaii under contract with the National Aeronautics and Space Administration.

## 3.2 Near-Infrared Observations of AS 205

We obtained JHK band observations of AS 205 on 1993 June 22 (UT) at the 3.8-meter United Kingdom Infrared Telescope (UKIRT) on Mauna Kea, Hawaii.[5] We used the facility infrared camera IRCAM, which employed a 58 × 62 pixel InSb hybrid array with a plate scale of 0.31″/pixel (McLean 1987). We flat-fielded with dark-subtracted twilight sky frames. Again, a flux standard close in airmass (S-R 3) was observed; its magnitudes were obtained from an unpublished UKIRT list, and absolute calibration was based on the work of Cohen *et al.* (1992). We estimated the image resolution in the same manner as for the mid-IR images. Table 2 summarizes our near-IR data, and Figure 2 shows our JHK images of AS 205.

Speckle observations of AS 205 were obtained on 1990 July 8, and 1992 May 14 (UT) using the Palomar Infrared Cassegrain Camera (*a.k.a.* D78) at the Hale 200-inch Telescope. The camera used a 58 × 62 pixel InSb hybrid array and had a plate scale of 0.053″/pixel when used with the f/415 Gregorian secondary. In July 1990, we observed 3 sets in H and K, and in May 1992, we observed 4 sets each at HKL′. The details of the observing processes are described in Ghez *et al.* (1993). To summarize, observations were taken in object/calibrator sets, each set consisting of 400 snapshots of 100 msec each of the object followed by 400 snapshots of a calibrator star. Two calibrators were observed to ensure that at least one was unresolved; SAO 159745 was the calibrator used for the reductions. Each snapshot image was sky-subtracted, flat-fielded, and corrected for bad pixels. Pairwise, the sets were analyzed to produce the object's diffraction-limited Fourier amplitudes and phases. The amplitudes and phases were combined in an inverse FFT to produce a diffraction-limited image.

## 3.3 Separating the Components of AS 205

We resolved both components of AS 205 in all our data. For our speckle data, the flux ratio and separation were most accurately determined by fitting the Fourier data as described by Ghez *et al.* (1995), where separation measurements for AS 205 were already reported. The final image was used only to remove the 180° position angle (p.a.) ambiguity in the power spectrum. Our results supercede the speckle results in Ghez *et al.* (1993). Unlike most binaries in their sample, AS 205's separation is a sizeable fraction of the Palomar IR camera's field of view; we eliminated poorly centered images from the data set and reduced the width of the tapered parameters, leading to more accurate results.

To measure the flux density ratios in our direct near and mid-IR images, we fit the two components using an analytic model for the point-spread function (psf). Our standard star images are not suitable for more sophisticated approaches, *e.g.*, psf-fitting or deconvolution, due to variable seeing conditions for our near-IR data and focus variations and IRTF image quality problems for our mid-IR data. To reduce systematic errors, we fit two models, an elliptical Gaussian and a two-dimensional Lorentzian, over a range of radii from the components. We averaged fitting results to determine the flux density ratios and used the standard deviation of the results to estimate $1\sigma$ random errors. We checked the fitting accuracy by repeating our procedure on simulated images of binary stars generated from the standard star images. The tests suggested a $\lesssim 10\%$ underestimate in our derived flux density ratios. However, at all wavelengths and over a range of relevant ratios, the systematic error between the fitted and true flux density ratio was within the estimated $1\sigma$ errors, confirming the accuracy of our fitting procedure. These tests also show that the astrometric results (§4.2) are accurate within the $1\sigma$ errors.

---

[5]UKIRT is operated by the Royal Observatory, Edinburgh, on behalf of the UK Science and Engineering Research Council.

# 4 Results

## 4.1 Spatial Extent of Mid−Infrared Emission

None of our images show new companions to our objects within the Berkeley camera's field of view, $3.9''$ east-west $\times$ $25''$ north-south. We tested our detection sensitivity to such companions by inserting artificial stars with a variety of flux ratios and separations into the images and then inspecting the images by eye. For each object, our tests were performed at the longest observed wavelength, where the resolution is coarsest. We find the data are sensitive to companions of equal brightness as close as the resolution limit given in Table 2. We are able to detect companions at most a factor of 10 fainter than the object at a separation of $2.0''$, a factor of 20 at $2.5''$, and a factor of 40 at $3''$, provided the companion flux is $\gtrsim 5\sigma$. After comparing the radial profiles of our objects with their corresponding standard stars, we also rule out any extended emission with significant azimuthal extent which is $\gtrsim 10\%$ of the peak flux. We conclude none of the stars in our sample are spatially extended within our field of view.

## 4.2 Resolving AS 205

Table 3 lists our derived flux density ratios for AS 205. Multiple measurements at the same wavelength are consistent with no changes in relative brightness except in the H band. Two-dimensional speckle observations by Zinnecker *et al.* (1990) find H and K flux density ratios comparable to our results and an L′ flux density ratio of $3.6 \pm 0.4$, somewhat larger than our data. In Figure 3, we show the SED for each component, constructed from our direct near-IR data and more recent mid-IR data, along with *IRAS* (Weaver & Jones 1992), sub-mm (Jensen, Mathieu, & Fuller 1995), and 1.3 mm (André & Montmerle 1994) fluxes for the total system.

Studies of PMS binaries have uncovered several "infrared companions," which are the optically fainter but infrared brighter star in these binaries (Mathieu 1994). Our IR data plus observations at Gunn-z (0.85–1.1 $\mu$m; Reipurth & Zinnecker 1993) and at HKL′ (Zinnecker *et al.* 1990) all find AS 205 North to be the brighter star. AS 205 South is the optically brighter star (Herbig & Rao 1972; Herbig & Bell 1988), suggesting AS 205 N is an IR companion. However, the non-simultaneity of the observations and the ambiguity in other optical data make this conclusion uncertain (§5.2.1).[6]

Table 4 lists the projected separation and p.a. from our data and from the literature. Adopting a distance of 160 pc to the Upper Scorpius association, AS 205 has a projected separation of 210 A.U. Our observations have insufficient time coverage to use the proper motion of stars in Upper Sco (Bertiau 1956) to exclude at the $2\sigma$ level the possibility that one of the stars is a distant background object. However, given the large K-brightnesses and similar physical properties of the components (§2.3), the binary is unlikely to be due to a chance projection. Ghez *et al.* (1995) discount the possibility that one of the stars is a Herbig-Haro object based on the small relative motion, and the low density of infrared-bright sources in Upper Sco makes it improbable the system is a chance projection of two T Tauri stars in the same cloud. We conclude AS 205 is a physically-bound pair.

---

[6]Based on the Herbig & Rao (1972) sighting, Herbig & Bell (1988) catalog the southern star as AS 205 (HBC 254) and the northern star as AS205/c (HBC 632).

## 4.3  Mid−Infrared Variability

We find all of our sources, with the possible exception of AS 205 S, are variable in the mid-IR. For most objects, we reach this conclusion by comparing our results with mid-IR photometry from the literature. The most reliable comparisons are those with the spectrophotometry ($\Delta\lambda/\lambda = 0.02$) of Cohen & Witteborn (1985), hereafter CW85, as opposed to using broad-band measurements, *e.g.*, *IRAS* 12 $\mu$m data ($\Delta\lambda/\lambda \approx 0.5$) where comparisons are complicated by the broad *IRAS* filters. Caution is also required when comparing 9.7 $\mu$m silicate feature data as it will be sensitive to small filter differences.

Our L1551-IRS5 12.0 $\mu$m flux from September 1990 is 90% larger than that measured by Cohen & Schwartz (1983) in December 1981. Since they used a larger aperture (11″) than our field of view, the flux difference suggests considerable mid-IR variability. Mid-IR photometry compiled by Molinari, Liseau, & Lorenzetti (1993) suggest N band (10.1 $\mu$m) variations with an amplitude of 0.4 mag, though the aperture sizes differ between observations.

Our September 1990 fluxes for HL Tau are consistent with the December 1982 data of CW85 and also with unpublished narrow-band mid-IR photometry from January 1981 at the Wyoming Infrared Observatory (Grasdalen 1995; Rydgren *et al.* 1984). However, Cohen & Schwartz (1976) observed a 0.9 mag increase in the N band flux over the span of only two nights in November 1974, indicating significant variability. Data taken at 11.1 $\mu$m over three nights in November 1973 by Rydgren, Strom, & Strom (1976) may also suggest variability.

Our May 1991 AS 209 fluxes at 9.7 and 12.5 $\mu$m are 30% less than CW85's May 1983 fluxes. Since CW85 used a 9″ beam, the difference might be due to a deeply embedded companion undetected at K-band (Ghez *et al.* 1993) which is beyond our east−west field of view. However, Cohen (1974), using an aperture (4″) comparable to our E−W field of view, find 8.5 $\mu$m and 12.5 $\mu$m fluxes approximately a factor of 3 and 2 greater than ours, suggesting AS 209 is quite mid-IR variable.

For AS 205, our May 1990 measurement of the total 12.5 $\mu$m flux is consistent with CW85; however, in May 1991, we measured a 50% increase at 12.5 $\mu$m along with a 40% increase at 8.5 $\mu$m and a possible 25% increase at 9.7 $\mu$m compared to CW85. Our May 1991 fluxes at 8.5 $\mu$m and 9.7 $\mu$m are also about 50% greater than those of Cohen (1974). Again, these comparisons are not exact due to the differing spectral resolutions, but clearly the system as a whole has become brighter in the mid-IR, particularly in the 8.5 and 12.5 $\mu$m continuum. It is unknown if these changes originated from one or both components. Based on our separation of the two components (§3.3), we do find AS 205 N brightened by 60% between May 1990 and May 1991 while AS 205 S remained unchanged, without any obvious correlation with the relative near-IR fluxes of the two stars over the same period of time (Table 3).

# 5 Discussion

## 5.1 Spatial Distribution of Mid-Infrared Emitting Material

The lack of spatially extended mid–IR emission in all of our sources is in accord with current understanding of the circumstellar material associated with embedded protostars and pre-main sequence stars. Nearly all the mid-IR flux for our objects originates from warm ($T \sim 300$ K) circumstellar dust heated directly or indirectly by radiation from the stellar photosphere and/or physical processes in the material, *e.g.*, viscous dissipation; mid-IR emission from the stellar photospheres is relatively weak. Interstellar grains are believed to be much smaller than our observing wavelengths, even for the large grains suspected to exist in dark star-forming clouds (Pendleton, Tielens, & Werner 1990); consequently, the effect of dust scattering is negligible (Spitzer 1978). The spatial distribution of the warm dust will depend on the evolutionary state of each YSO, but in general the warm material will be only few A.U. from the central source. For instance, a blackbody grain 1 A.U. from a 1 $L_\odot$ star has an equilibrium temperature of 280 K and emits most strongly at 10 $\mu$m.

The emission from the protostar L1551-IRS5 has been modelled in detail (Adams, Lada, & Shu 1987; Butner *et al.* 1991; Kenyon, Calvet, & Hartmann 1993; Butner, Natta, & Evans 1994). All these models agree with the lack of spatial extent in our images.

"Flat spectrum" sources such as HL Tau, *i.e.*, those with flat SEDs, likely have emission from both a disk and from residual infalling material (Calvet *et al.* 1994), making these objects transition cases from Class I to Class II (Lada 1991). The nominal model of Calvet *et al.* has a characteristic mid-IR size of 1–2 A.U. so we expect HL Tau to be unresolved, as we have observed. Note that future mid-IR astrometry would be useful to settle the issue of the location of the central star (Stapelfeldt *et al.* 1995; Weintraub *et al.* 1995); our data do not cover a sufficient field of view to address this point.

The CTTS-type optical spectra and IR excesses of AS 209 and both components of AS 205 suggest they possess circumstellar disks (*e.g.*, Strom 1994). AS 209's SED[7] begins to rise at 100 $\mu$m, and the object has a large 1.3 mm flux (André & Montmerle 1994); both facts imply large quantities of cold dust. The non-photospheric 2–60 $\mu$m emission is roughly a power-law for AS 209 and both stars in AS 205 (§5.2.2), suggesting it arises from circumstellar disks with power-law temperature distributions (Lada 1991). These disks are expected to have characteristic mid-IR sizes of only a few A.U. and are therefore expected to be unresolved (Adams *et al.* 1987; Kenyon & Hartmann 1987).

Cohen & Witteborn (1985) speculated that the 9.7 $\mu$m silicate feature might be used to infer the viewing geometry to a YSO, with an emission feature being produced by a more pole-on line of sight to a circumstellar disk and an absorption feature by a more edge-on viewing angle. This feature is seen in emission in AS 209 and both components of AS 205 and in absorption in L1551-IRS5 and HL Tau (Cohen & Schwartz 1983; CW85). In fact, a variety of geometries can produce the feature in emission as there may be optically thin and thick components below our resolution limit. A temperature inversion in a circumstellar structure can also produce an emission feature (*e.g.*, Calvet *et al.* 1991). While the 9.7 $\mu$m feature is a useful observational constraint for theoretical models, its presence alone does not reveal the viewing geometry.

---

[7] We constructed the SED of AS 209 using photometry from the literature (Herbig & Bell 1988; Glass & Penston 1974; Weaver & Jones 1992) and estimated the photospheric SED as in §5.2.1 using the published spectral type and $A_V$ (Basri & Bathala 1990; Cohen & Kuhi 1979; *cf.* Valenti *et al.* 1993).

## 5.2 Components of AS 205

### 5.2.1 Which Star is Which?

Our 1–12 $\mu$m data resolve both components of AS 205, allowing us to isolate the IR emission from each. Our goal is to use the SEDs to probe the stars' evolutionary states and circumstellar material so we require a spectral type and extinction for each star to account for the photospheric flux. Unfortunately, this information is not well-determined. Cohen & Kuhi (1979), hereafter CK79, performed the only spectral type and $A_V$ measurements which resolved both stars. They found the brighter star had a spectral type of K0, and the other star had a K5 spectral type in May 1975, but they did not record the p.a. of the binary so we must infer which star was optically brighter at the time to use their results. Such an attempt is complicated by the system's optical (Herbig & Rao 1972; Herbig & Bell 1988; Covino *et al.* 1992; Basri & Stassun 1994) and near-IR (Glass & Penston 1974; Cohen 1975; Whitelock & Feast 1984) variability as well as possible spectroscopic variations (Herbig 1977; Herbig & Bell 1988; and references in §2.3).

Figure 4 plots the SEDs of the two stars using our IR data and the V data point of CK79 along with the flux from the underlying photospheres in the two possible cases. We have used the CK79 spectral types and $A_V$'s with an ordinary extinction law (Mathis 1990),[8] assumed that T Tauri photospheres have the colors of main sequence dwarfs (Bessel & Brett 1988; Hartigan *et al.* 1994), and assumed the J band flux is entirely from the photosphere to minimize contamination from any disk emission. In Figure 4a, we assumed AS 205 N was the optically brighter K0 star seen by CK79 and AS 205 S the dimmer K5 star; the resulting V band fit to the photospheres is quite poor. Figure 4b shows the alternative case, where AS 205 S is the optically brighter K0 star. The fit looks quite good; however, it disagrees with the observation that the northern star is 5 times brighter at ∼0.9 $\mu$m than the southern star (Reipurth & Zinnecker 1993). We can conceive of scenarios which employ plausible variability effects and the non-simultaneity of the measurements to improve the fits. However, *none of the possible explanations can uniquely identify the spectral types of AS 205 N and AS 205 S*. Future spectroscopic observations which resolve both stars are needed to establish defintive optical and spectral type identifications.

### 5.2.2 Circumstellar Environments

Despite the identification ambiguity just discussed, we are able study the circumstellar environments of both components of AS 205. As shown in Figure 4, both stars exhibit IR excesses. The SEDs of both stars resemble those of Class II YSOs, with the spectrum of AS 205 S slightly steeper. The 9.7 $\mu$m emission feature is present in both sources, though more pronounced in AS 205 S.

---

[8]CK79 determined $A_V = 3.0 \pm 0.2$ for the brighter K0 star and $A_V = 5.8 \pm 0.3$ for the fainter K5 star using narrow-band colors from 5400–6700 Å. It is plausible the K5 star in fact had a somewhat later spectral type because of contamination in its spectrum from the brighter K0 star. Using CK79's original data and spectral standards, we recalculated the visual extinction to the fainter star assuming a spectral type of M0 and found $A_V = 5.2 \pm 0.3$. The subsequent change in the J band extinction, which constrains our photospheric fit, is negligible. Therefore, the possible error in the fainter star's spectral type will not qualitatively change our subsequent discussion, although the quantitative details would be slightly altered.

If we attribute all the IR excesses to dust in disks (§5.1), we can estimate the disk optical depth in the mid-IR (Skrutskie *et al.* 1990). A disk will intercept some of the stellar light and reprocess it to longer wavelengths. The mid-IR reprocessing flux can be written as $F_\lambda^{phot} \times 10^{(\Delta N/2.5)}$, where $F_\lambda^{phot}$ is the photospheric flux and $\Delta N$ is the excess $N$ band (10.1 $\mu$m) emission in mags due to reprocessing. To calculate the expected reprocessing flux, we find $F_\lambda^{phot}$ as in §5.2.1 and use $\Delta N$ as calculated by Hillenbrand *et al.* (1992) for a spatially thin, optically thick disk viewed pole-on. For such a disk around a K0 star, $\Delta N = 3.4$ mag and around a K5 star, $\Delta N = 3.1$ mag. The choice of pole-on geometry gives us upper limits on $\Delta N$ from a flat reprocessing disk; however, any material in any extended envelope (§5.3) will increase the mid-IR reprocessing flux and may cause us to overestimate the disks' optical depths. We averaged our May 1991 photometry to approximate the $N$ band fluxes. If AS 205 N is the K0 star and AS 205 S is the K5 star, we find $\Delta N = 5.2 \pm 0.5$ mag for AS 205 N and $\Delta N = 3.2 \pm 0.8$ mag for AS 205 S. In the reverse case, we find $\Delta N = 4.2 \pm 0.5$ mag for the K5-type AS 205 N and $\Delta N = 4.4 \pm 0.8$ mag for the K0-type AS 205 S.[9] If we use our May 1990 data, $\Delta N$ is lowered by 0.5 mag for AS 205 N. In all these cases, the earlier spectral type star has a larger mid-IR excess. Since the observed mid-IR excesses exceed those possible from reprocessing alone, we conclude that both circumstellar disks are optically thick in the mid-IR and that there must also be an additional source of mid-IR flux, *e.g.*, accretion.

### 5.2.3 Evolutionary Status

In Figure 5, we explore the evolutionary status of both components of AS 205 by comparing their positions on the HR diagram with theoretical predictions by D'Antona & Mazzitelli (1994) which use the convection recipe of Canuto & Mazzitelli (1992) and the opacities of Alexander, Augason, & Johnson (1989). To place the stars on the horizontal axis of the HR diagram, we choose AS 205 N to be the K0 star and AS 205 S to be the K5 star and use the Hartigan *et al.* (1994) effective temperature scale for main-sequence dwarfs. We adopt an error estimate of 2 subclasses in the spectral type of AS 205 N (CK79) and consider the position of AS 205 for spectral types of K5 and M0 to account for contamination in the spectral typing by the brighter K0 star (§5.2.1). Using J band bolometric corrections from Hartigan *et al.* (1994) and a distance modulus of $6.0 \pm 0.8$ (160 pc; de Geus *et al.* 1989), we compute the bolometric luminosity $L_*$ of each underlying star: $(L_*)_N = 8.2 \pm 1.3 \, L_\odot$, $(L_*)_S = 4.9 \pm 1.7 \, L_\odot$. From Figure 5, we find masses of 2.2 $M_\odot$ and 0.7 $M_\odot$ and approximate ages of $10^6$ and $2 \times 10^5$ yr for AS 205 N and AS 205 S, respectively. Use of different opacities and convection models in the D'Antona & Mazzitelli models and also use of models by Swenson *et al.* (1994) all lead to similiar age discrepancies; the models suggest the AS 205 system is non-coeval by at least a factor of 3. Note that if we had chosen AS 205 S as the K0 star and AS 205 N as the K5 star, the mass and age differences would have been even larger. Our results are in accord with Hartigan *et al.* (1994) who find for widely separated binaries (400–6000 A.U.) in non-coeval systems that the less massive star always seems

---

[9]To determine the 1$\sigma$ random errors in $\Delta N$, we summed in quadrature errors from the J flux density ratio (§3.3), $A_V$ (CK79), possible boundary layer emission (Hartigan *et al.* 1994), and variability. We estimated the J band variability amplitude to be 0.2 mag and assumed a 0.14 mag variability for each component (§5.3). J band variability dominates the error budget for AS 205 N and is comparable to the error from the flux density ratio for the error budget of AS 205 S.

younger.

Systematic uncertainties in $L_*$ can arise from three sources: the distance to AS 205, undetected companions, and accretion effects. Although any distance error would not affect the flux density ratio since the two stars are a physical pair, it would mildly affect the mass and age ratios since the convective tracks are not strictly vertical and the isochrones are not evenly spaced. It is unlikely our results for $L_*$ are affected by undetected companions. Ghez *et al.* (1993) establish a minimum K flux density ratio of $7 \pm 1$ for an additional companion with a separation of 0.10–0.80″; such a companion would introduce a systematic error in the derived $L_*$ comparable to the $1\sigma$ error for AS 205 N and smaller than the $1\sigma$ error for AS 205 S. Accretion effects could be the most serious source of error. Since we have assumed the J band flux is entirely photospheric, any J band accretion luminosity will lead to an overestimate of $L_*$ and consequently an underestimate of the age.[10] In addition, if the mass accretion rate is quite large, progress along the evolutionary tracks will be slowed, again making older stars appear younger (Hartmann & Kenyon 1990). The magnitude of these effects could also differ for each component. Nevertheless, the strong IR and mm emission from the AS 205 suggests large quantities of circumstellar material and attest to the system's youth.

Theoretical bias toward coeval formation of binaries as opposed to non-coeval binaries formed by capture (*e.g.*, Pringle 1991; Ostriker 1994) might imply some mechanism causes the apparent non-coevality of AS 205. But the fact is that reliable mass and age estimates for PMS stars cannot be derived from current theoretical models, because, for instance, accretion effects could be quite severe. More importantly, there has yet to be an independent check of the models. Testing the models with mass determinations will soon be possible for very close PMS binaries, but such mesaurements for wider separation binaries like AS 205 will take many decades.

## 5.3  Nature of the Mid-Infrared Variability

The magnitude and timescales of mid-IR variability provide insights into the physical processes occuring in the circumstellar material. Mid-IR variability in YSOs has been virtually unexplored, aside from studies of T Tau (Ghez *et al.* 1991) and a few exciting sources of Herbig-Haro objects (Molinari *et al.* 1993; Molinari 1990). Unlike the optical and near-IR emission from YSOs, nearly all the mid-IR flux originates from the circumstellar material; the cause of the mid-IR variability may originate there as well. As our data are sparsely sampled in time and spectral range, we limit our analysis to an exploration of possible causes. We postulate a plausible origin for the variations is time-varying mass accretion rates in the disks of these objects.

The younger objects in our sample, L1551-IRS5 and HL Tau, are believed to be obscured by dusty envelopes of infalling material. Any time-varying luminous accretion events (*e.g.*, Bell, Lin, & Ruden 1991) will be reprocessed by the envelope, resulting in time-varying mid-IR fluxes. Such processes are suspected to power FU Ori objects, and L1551-IRS5 is probably a low-luminosity version of these objects (Mundt *et al.* 1985; Stocke *et al.* 1988; Carr, Harvey, & Lester 1987). In addition, even the youngest YSOs are thought to have circumstellar disks, and instabilities in the disks could easily vary the mid-IR flux, *e.g.*, if a density perturbation changes the disk temperature structure (Shu *et al.* 1990). Another possibility is that the mid-IR variability is related to the outflow activity seen in both L1551-IRS5 and HL Tau, *e.g.*, through the interaction of the outflow with the dusty envelope.

---

[10]In fact, there is an indication we have overestimated $L_*$. The integrated luminosity of AS 205 (derived from non-simultaneous data) is about 11 $L_\odot$ whereas the sum of our calculated stellar bolometric luminosities is 12.7 $L_\odot$. Our J band flux, which we have assumed is entirely photospheric, is 0.4–0.8 mag higher than past measurements (§5.2.1). However, we use our observation as it is the only one with photometry and which resolves both stars.

For the more evolved CTTS like AS 205 N, AS 205 S, and AS 209, the circumstellar material likely resides in a disk. Viscous accretion disk theory has proven useful in explaining the IR emission from these objects (Adams *et al.* 1987; Kenyon & Hartmann 1987). In such an interpretation, the steady-state mid-IR emission can be partitioned into (1) the photospheric flux; (2) the stellar flux absorbed and reprocessed; and (3) the flux intrinsic to the disk. The photospheric contribution (1) is negligible compared to (2) and (3). Variations in the reprocessing flux (2) will occur if the direct radiation from the stellar photosphere changes in brightness. Photometric monitoring of CTTS from U band to the near-IR find periodic and quasi-periodic fluctuations, attributed to large hot and cool spots on the stellar surface (*e.g.*, Kenyon *et al.* 1994; Guenther & Hessman 1993). The spots are inferred to be ~1000 K cooler or warmer than the stellar photosphere and to have covering fractions of 1–20% (Bouvier & Bertout 1989; Bouvier *et al.* 1993). As in §5.2.2, we can calculate the reprocessing flux from a geometrically thin, optically thick disk viewed pole-on. For AS 205 North and AS 209, in the extreme case of 100% spot coverage we find disk reprocessing could generate at most 15% changes in the flux. If the disk is flared at large radii, the mid-IR contribution from reprocessed light might be enhanced by a factor of 2–3 (Kenyon & Hartmann 1987; *cf.* Adams *et al.* 1988). We conclude variations in the reprocessing flux can account for only minor perturbations in the mid-IR emission.

Changes in the mass accretion rate through the disk are the most likely source of the mid-IR variability of AS 205 N and AS 209. For a geometrically thin, optically thick accretion disk in thermal equilibrium, the radial temperature profile depends on the distance $R$ from the star and the mass accretion rate $\dot{M}$ as

$$T_{acc}(R) \sim R^{-3/4} \dot{M}^{1/4}$$

(Shakura & Sunyaev 1973; Lynden-Bell & Pringle 1974). Approximating the emission from each disk annulus as being emitted at the blackbody curve peak, the monochromatic accretion luminosity is

$$(\lambda F_\lambda)_{acc} \sim \lambda^{-4/3} \dot{M}^{2/3}.$$

Note that changes in $\dot{M}$ do not alter the spectral slope, only the emission amplitude. After subtracting the photospheric and reprocessing flux, we can convert the mid-IR variations to changes in $\dot{M}$. For AS 205 North, the difference between the May 1990 and 1991 data suggest a factor 3 increase in $\dot{M}$. For AS 209, we find the factor 2–3 change in mid-IR flux implies a factor of 3–7 change in $\dot{M}$. Our assumption of pole-on viewing geometry means these numbers are lower bounds on the mass accretion rate variability.

We have not considered any mid-IR emission from material not in a disk, *e.g.*, from an envelope formed by a wind from the disk surface (Safier 1993; Natta 1993) or from residual infalling material (Calvet *et al.* 1994). Such mechanisms have been used to explain why CTTS SEDs do not follow the slope expected from classical viscous accretion disks, *i.e.*, $\lambda F_\lambda \sim \lambda^{-4/3}$. If the mid-IR contribution from these non-disk sources is relatively steady, the variations in $\dot{M}$ would be much larger than our estimates.

Investigation of variability mechanisms beyond the simple considerations discussed here will require considerably more observations. Repeated mid-IR measurements are needed to measure the variability timescale and to constrain the origin of this phenomenon. Short timescales, *i.e.*, hours or days, would point to processes near the surface of the star, *e.g.*, in the star-disk interface, as the culprit of the mid-IR variations. Longer term variations, *i.e.*, months or years, might suggest processes in the circumstellar material, *e.g.*, variations in the accretion rate as we have suggested. Simultaneous multiwavelength monitoring will offer even more gains. The optical and/or near-IR variability in some YSOs is known to be correlated (*e.g.*, Kenyon *et al.* 1994; Liseau, Lorenzetti, & Molinari 1992); any connection to the mid-IR variability is currently unknown. YSOs may also be variable at even longer wavelengths; this would certainly be the case if the mass accretion rate through the disk is time-varying.

Past modelling of the SEDs of YSOs has invariably depended on non-simultaneous photometry from optical to millimeter wavelengths (*e.g.*, Adams *et al.* 1987; Hillenbrand *et al.* 1992; Kenyon *et al.* 1993). The mid-IR variability argues accurate models of the circumstellar material cannot be derived by using a melange of published data. The measured slope of the non-photospheric SED can vary considerably depending on the choice of non-simultaneous data. For instance, the measured slope of AS 205 N's non-photospheric emission, *i.e.*, $\lambda F_\lambda \sim \lambda^{-\alpha}$, changes from $\alpha = 0.3$ to $\alpha = 0.0$ if we use our 1993 K band photometry with our 12.5 $\mu$m data points from 1990 and 1991. For AS 209 the slope can vary from $\alpha = 0.0$ (using our mid-IR data) to $\alpha = 0.8$ (using mid-IR data from Cohen 1974). Clearly, the physical interpretation of these sources will differ with these two slopes.

# 6 Conclusions

We have presented arcsecond resolution mid-IR (8–13 $\mu$m) imaging data of four YSOs, L1551-IRS5, HL Tau, AS 205, and AS 209. We find no evidence for spatially extended emission with significant azimuthal extent within our field of view and sensitivity limits. The lack of extended emission is consistent with current understanding of the circumstellar material of YSOs. We also find no new companions. We have verified the single stars in our sample do not have any mid-IR emission from previously undetected cool companions and have resolved the components of the close binary AS 205 in the mid-IR for the first time. We have also presented near-IR speckle and direct imaging data of AS 205 which shows both components.

We find AS 205 North to be brighter and to have a flatter SED from 1–12 $\mu$m than AS 205 South. On the other hand, visual data find AS 205 S to be brighter so AS 205 N may be an example of an infrared companion, though this conclusion is tentative since the data are non-simultaneous. The published optical spectroscopy is insufficient to assign spectral types to the individual components, but we are still able to draw several conclusions. If the stars' IR excesses arise from geometrically thin circumstellar disks, the SEDs suggest the relevant emitting regions (the inner few A.U.) are optically thick in the mid-IR. In contrast, binaries with smaller separations show evidence of inner disks optically thin in the mid-IR. Furthermore, the IR excesses of AS 205 N and AS 205 S are too large to originate solely from disk reprocessing of stellar radiation; there must be an additional contribution from, for example, accretion. Pre-main sequence evolutionary models suggest the binary is non-coeval, but the validity of these models is not currently established. The theoretically derived ages are $10^5$–$10^6$ yr. The bright mid-IR, sub-mm, and mm emission also suggests AS 205 is reasonably youthful. Future mid-IR surveys of a large number of YSOs will be important in exploring whether binaries can form in which one star is very deeply embedded while the other is optically revealed; the discovery of such systems might suggest binaries can form non-coevally.

All of our objects, with the possible exception of AS 205 S, exhibit mid-IR variability. In the most extreme case, AS 209, the mid-IR flux has changed by a factor of 3. For the classical T Tauri stars AS 205 N and AS 209, variations in the direct heating of their circumstellar disks by accretion-related photospheric hot and cool spots may explain the optical and near-IR variability, but it cannot account for the magnitude of the mid-IR variability. We suggest the mid-IR variability reflects fluctuations in the mass accretion rate $\dot{M}$ through the circumstellar disk. The implied changes in $\dot{M}$ are about an order of magnitude. The mid-IR variability implies that past modelling of the circumstellar material of YSOs may have been flawed because of non-simultaneous photometry. Future campaigns which acquire simultaneous multiwavelength data with good temporal resolution will provide valuable insights into the physical processes in the circumstellar material of young stars.

We are grateful to Martin Cohen, George Herbig, Lynne Hillenbrand, Joan Najita, and Matt Richter for helpful discussions. We thank Eugenia Katsigris for some of the data reductions and Karen Strom and the Department of Physics and Astronomy at the University of Massachusetts for making the pre-main sequence evolutionary models available on the World Wide Web. It is a pleasure to acknowledge Matt Richter and Dan Stevens for a close reading of this manuscript. This research has made use of the Simbad database, operated at CDS, Strasbourg, France. During the course of this work MCL has been supported by an NSF Graduate Student Fellowship and CalSpace grant CS-70-93 and JRG by a fellowship from the Packard Foundation.

Table 1: OBJECT LIST

| Object | HBC[a] # | RA (1950) | Dec (1950) | $m_V$ | $m_K$ | Sp. Type | $d$ (pc) | Location |
|--------|----------|-----------|------------|-------|-------|----------|----------|----------|
| L1551-IRS5 | 393 | 04 28 40 | +18 01 41 | — | 9.1[b] | K2III? | 140[c] | Tau-Aur, L1551 cloud |
| HL Tau | 49 | 04 28 44 | +18 07 36 | 14.6 | 7.1 | K7/M2? | 140[c] | Tau-Aur, L1551 cloud |
| AS 205 N+S | 632+254[d] | 16 08 38 | −18 30 43 | 12.4 | 5.8 | K5+K0[d] | 160[e] | Upper Sco, B40 cloud |
| AS 209 | 270 | 16 46 25 | −14 16 56 | 11.5 | 6.8 | K5 | 125[f] | east of $\rho$ Oph cloud |

[a] Herbig & Bell 1988

[b] Leinert *et al.* 1993

[c] Elias 1978

[d] see §5.2.1

[e] Jones 1970; de Geus, de Zeeuw, & Lub 1989

[f] distance to center of $\rho$ Oph cloud from de Geus *et al.* 1989

Table 2: DIRECT IMAGING OBSERVATIONS AND PHOTOMETRY

| Object | $\lambda$ | Date | Standard | $t_{on-source}$ | Resolution | $\sigma^a$ | $F_\nu$ |
|---|---|---|---|---|---|---|---|
| | ($\mu$m) | (UT) | | (sec) | ($''$ FWHM) | (Jy/arcsec$^2$) | (Jy) |
| L1551-IRS5 | 12.0 | 1990 Sep 27 | $\alpha$ Tau | 100 | 1.2 | 0.04 | $9.1 \pm 0.9$ |
| HL Tau | 8.5 | 1990 Sep 27 | $\alpha$ Tau | 125 | 1.1 | 0.04 | $5.3 \pm 0.5$ |
| | 9.7 | | | 62 | 1.1 | 0.06 | $3.8 \pm 0.4$ |
| | 12.0 | | | 88 | 1.2 | 0.04 | $6.2 \pm 0.6$ |
| AS 205 | 1.2 (J) | 1993 Jun 22 | S-R 3 | 29 | 1.5 | $1.9 \times 10^{-4}$ | $1.26 \pm 0.05$ |
| | 1.6 (H) | | | 29 | 1.5 | $2.2 \times 10^{-4}$ | $2.17 \pm 0.08$ |
| | 2.2 (K) | | | 29 | 1.4 | $2.4 \times 10^{-4}$ | $3.9 \pm 0.1$ |
| | 8.5 | 1991 May 14 | $\alpha$ Her | 103 | 1.1 | 0.03 | $8.2 \pm 0.8$ |
| | 9.7 | | | 98 | 1.2 | 0.04 | $10.7 \pm 1.1$ |
| | 12.5 | | | 98 | 1.2 | 0.04 | $11.4 \pm 1.1$ |
| | 12.5 | 1990 May 11 | $\phi$ Oph | 66 | 1.2 | 0.04 | $7.2 \pm 0.7$ |
| AS 209 | 8.5 | 1991 May 15 | $\alpha$ Her | 94 | 1.1 | 0.05 | $1.2 \pm 0.1$ |
| | 9.7 | | | 112 | 1.0 | 0.04 | $2.2 \pm 0.2$ |
| | 12.5 | | | 141 | 1.2 | 0.04 | $2.0 \pm 0.2$ |

[a] $1\sigma$ noise in the final mosaics (see §3)

Table 3: AS 205 FLUX DENSITY RATIOS

| $\lambda$ ($\mu$m) | Date (UT) | Observing Mode | $F_\lambda(North)/F_\lambda(South)$ |
|---|---|---|---|
| 1.2 (J) | 1993 Jun 22 | direct | $2.7 \pm 0.3$ |
| 1.6 (H) | 1993 Jun 22 | direct | $2.1 \pm 0.5$ |
| 1.6 (H) | 1992 May 14 | speckle | $2.4 \pm 0.1$ |
| 1.6 (H) | 1990 Jul 08 | speckle | $2.0 \pm 0.1$ |
| 2.2 (K) | 1993 Jun 22 | direct | $2.4 \pm 0.2$ |
| 2.2 (K) | 1992 May 14 | speckle | $2.49 \pm 0.04$ |
| 2.2 (K) | 1990 Jul 08 | speckle | $2.7 \pm 0.2$ |
| 3.8 (L$'$) | 1992 May 14 | speckle | $2.7 \pm 0.1$ |
| 8.5 | 1991 May 14 | direct | $5.7 \pm 0.5$ |
| 9.7 | 1991 May 14 | direct | $3.7 \pm 0.5$ |
| 12.5 | 1991 May 14 | direct | $6.5 \pm 1.9$ |
| 12.5 | 1990 May 11 | direct | $4.4 \pm 0.4$ |

Table 4: AS 205 BINARY PARAMETERS

| Date (UT) | Separation ($''$) | p.a. ($°$) | $\lambda\lambda$ | Reference |
|---|---|---|---|---|
| 1971 Jun 29 | $\sim$1 | $\sim$20 | optical | Herbig & Rao 1972 |
| 1975 May 31 | $\sim$1 | ?? | optical | Cohen & Kuhi 1979 |
| 1986 Jun 17 | $\sim$1.4 | $\sim$20/200 | optical | Herbig & Bell 1988; Herbig 1994 |
| 1986–1992 | 1.4 $\pm$ ?? | 204 $\pm$ ?? | near-IR | Reipurth & Zinnecker 1993 |
| 1990?? | 1.45 $\pm$ 0.03 | 211 $\pm$ 4 | near-IR | Zinnecker *et al.* 1990 |
| 1990 May 11 | 1.46 $\pm$ 0.09 | 212.1 $\pm$ 5.4 | mid-IR | this work |
| 1990 Jul 08 | 1.31 $\pm$ 0.03 | 212.0 $\pm$ 1.0 | near-IR | Ghez *et al.* 1995 |
| 1991 May 14 | 1.35 $\pm$ 0.06 | 213.5 $\pm$ 3.4 | mid-IR | this work |
| 1992 May 14 | 1.32 $\pm$ 0.03 | 212.0 $\pm$ 1.0 | near-IR | Ghez *et al.* 1995 |
| 1993 Jun 22 | 1.33 $\pm$ 0.07 | 215.7 $\pm$ 4.2 | near-IR | this work |

Note: The position angle (p.a.) is defined as the angle between the binary axis and North where the brighter star at the particular observing wavelength is at the origin. See §5.2.1 for a discussion of the optical data. The question marks indicate the information is not available in the literature.

# Figure Captions

FIGURE 1: Mid-IR images of AS 205 in May 1991. Contours start from $3\sigma$ and are spaced by a factor of 2. The lowest contour levels are 0.09 Jy/arcsec$^2$, 0.12 Jy/arcsec$^2$, and 0.12 Jy/arcsec$^2$ at 8.5, 9.7, and 12.5 $\mu$m, respectively.

FIGURE 2: JHK images of AS 205, contoured as in Figure 1. The lowest contour levels are 51 mJy/arcsec$^2$, 66 mJy/arcsec$^2$, and 72 mJy/arcsec$^2$ at JHK, respectively.

FIGURE 3: Spectral energy distributions for AS 205 North and South constructed from our near-IR (JHK) and more recent mid-IR (8.5, 9.7, 12.5 $\mu$m) data. The astericks are flux measurements for the combined system at the *IRAS* 25, 60, and 100 $\mu$m bandpasses (Weaver & Jones 1992), sub-mm/mm wavelengths (350, 450, 800, 1100 $\mu$m; Jensen *et al.* 1995), and 1.3 mm (André & Montmerle 1994). The $1\sigma$ errors are comparable to or smaller than the size of the plotting symbols, except for the 100 $\mu$m data point.

FIGURE 4: The observed SEDs of AS 205 North and South along with V band photometry from Cohen & Kuhi (1979). The lines (*dashed* for AS 205 N, *dotted* for AS 205 S) represent reddened photospheres for the underlying stars, constrained to fit the J band fluxes (§5.2.1). In (a), we have assumed AS 205 N was the optically brighter K0 star and AS 205 S was the optically fainter K5 star seen by Cohen & Kuhi. In (b) we have chosen the reverse case. Notice the components' mid-IR excesses are quite large in either case, in fact too large to be attributed to only disk reprocessing of stellar flux. There must be an additional mid-IR emitting source, *e.g.*, accretion luminosity (§5.2.2).

FIGURE 5: The positions of AS 205 N and S on the HR diagram juxtaposed against the evolutionary models of D'Antona & Mazzitelli (1994) using the opacities of Alexander *et al.* (1989) and the convection model of Canuto & Mazzitelli (1992). The dotted lines represent the evolution of stars of fixed mass. The solid lines are isochrones for $10^5$, $3\times10^5$, $10^6$, $3\times10^6$, and $10^7$ yrs (from right to left). The vertical error bars show the $2\sigma$ errors in $L_*$. For AS 205 N we have adopted a spectral type of K0 and an error of two subclasses. We show the position of AS 205 S for spectral types of K5 and M0 to account for possible contamination from AS 205 N (§5.2.1). The two stars seem to be non-coeval, a result which does not depend on the choice of evolutionary model (§5.2.3).

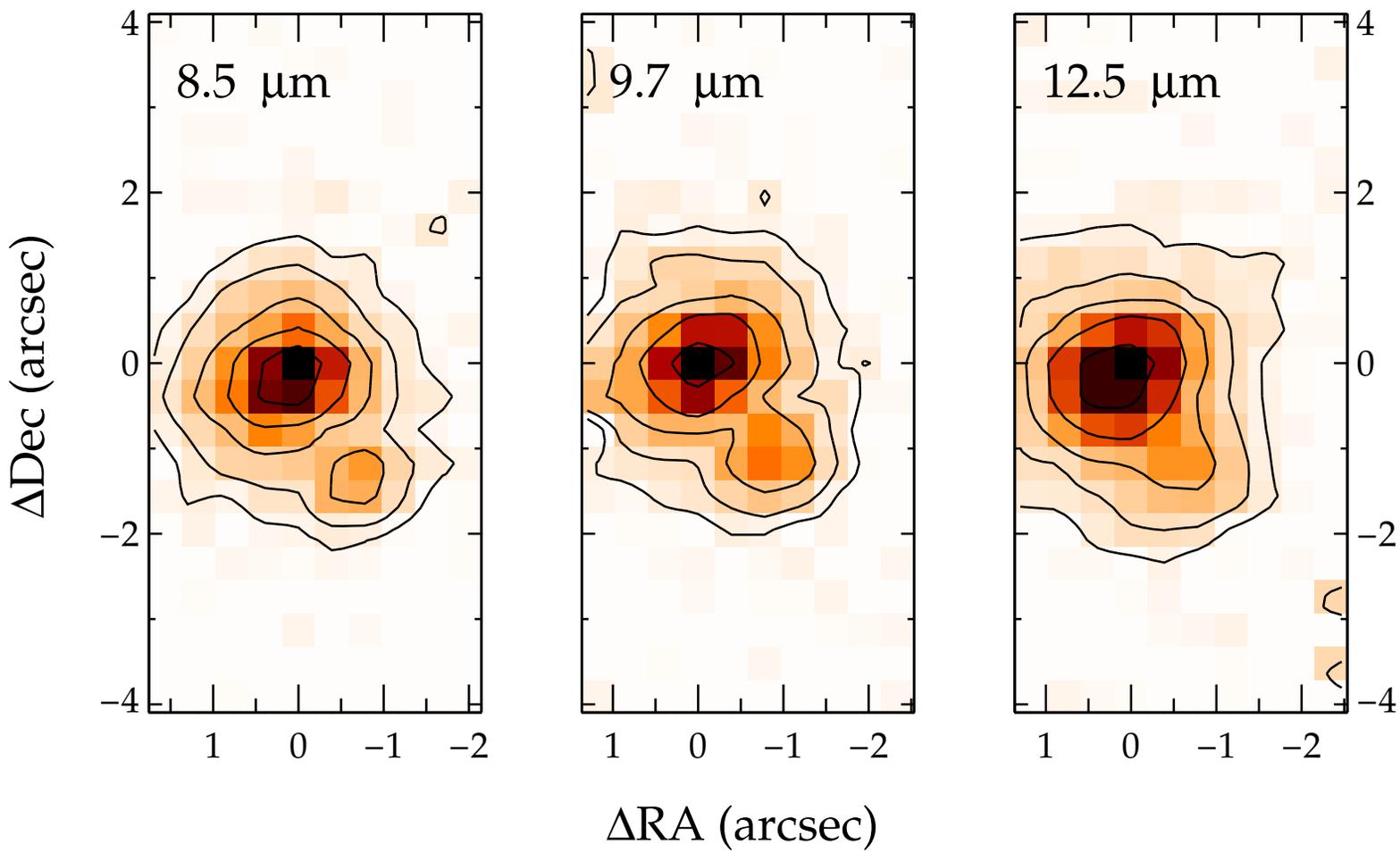

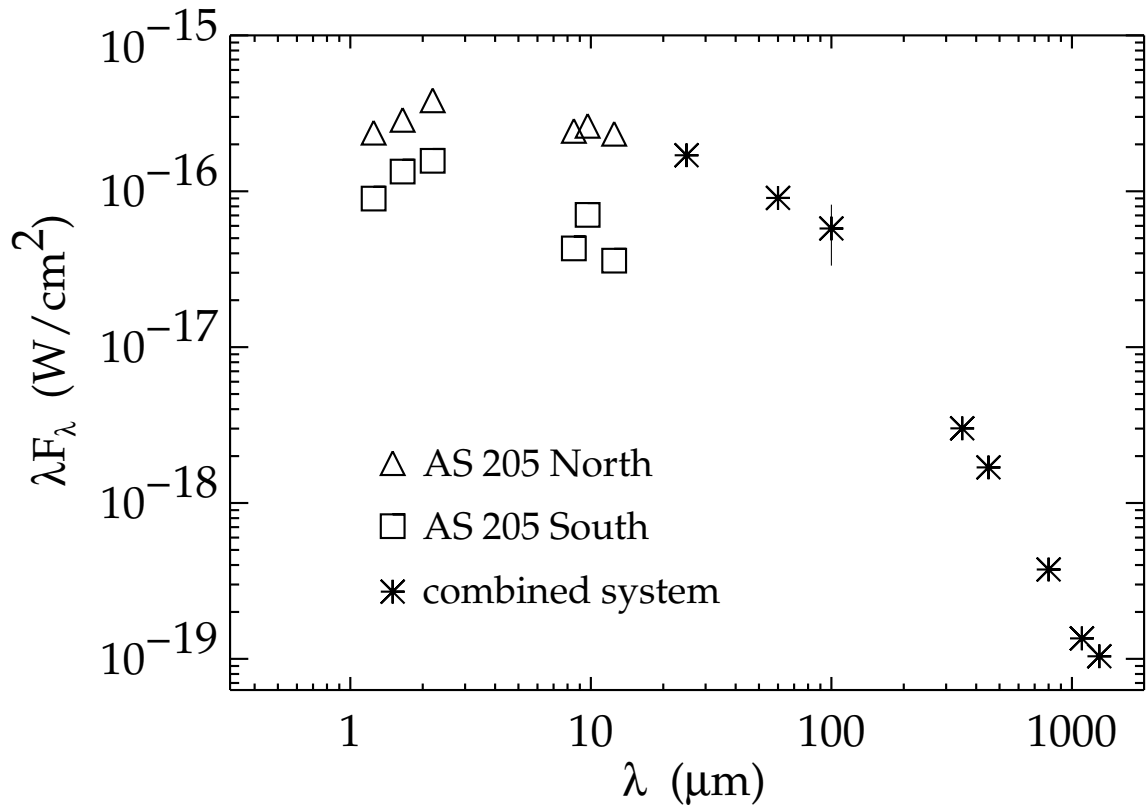

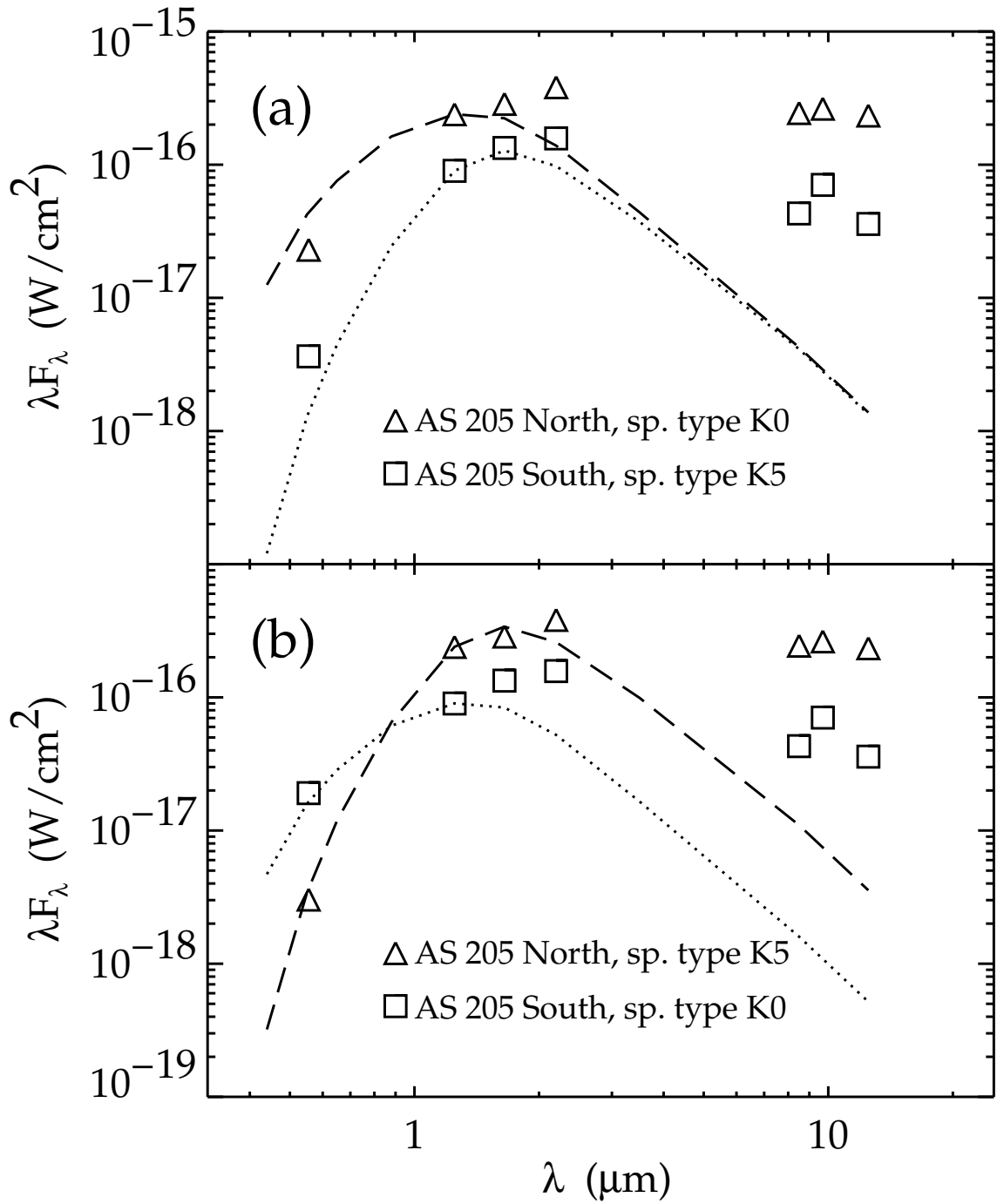

(a)

AS 205 North, sp. type K0
AS 205 South, sp. type K5

(b)

AS 205 North, sp. type K5
AS 205 South, sp. type K0

FIGURE 4

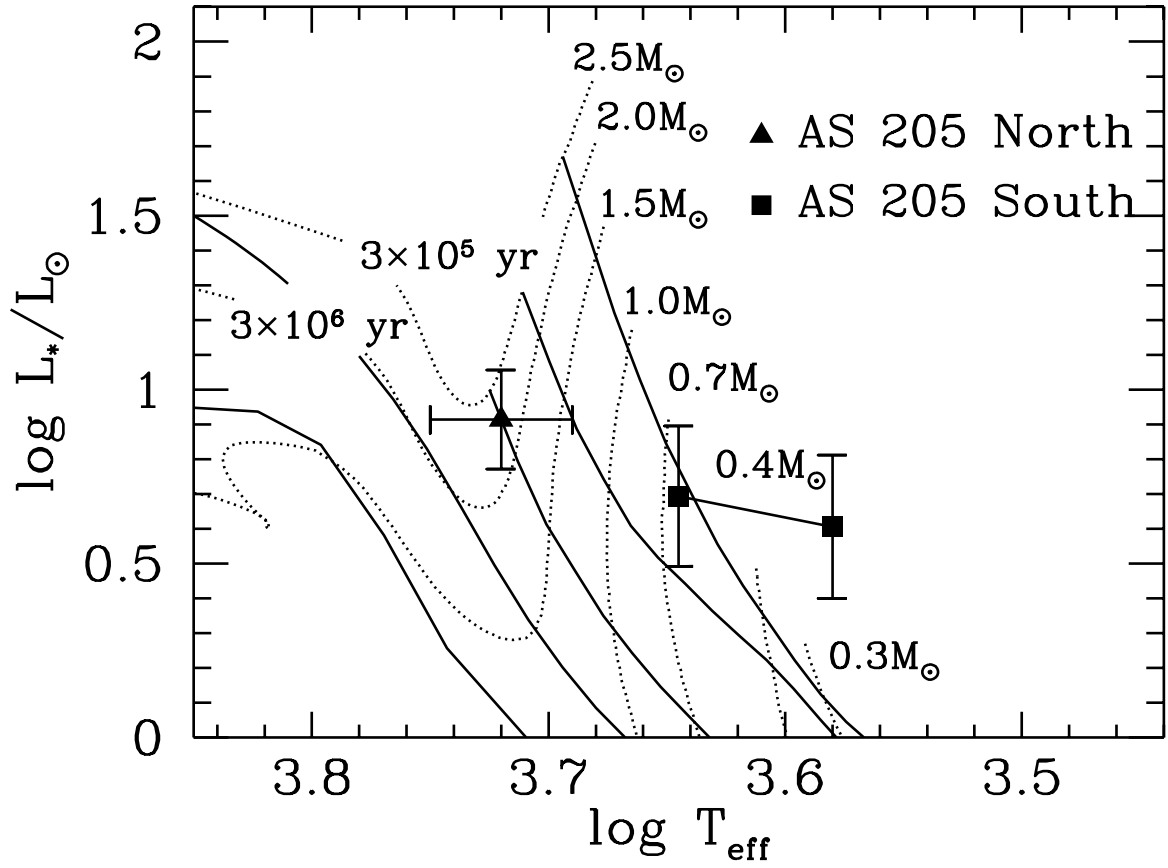